\documentclass[camera,letterpaper,nomarginnotes,nonarrowgutter]{jpaper}

\usepackage{amsmath,amssymb,amsfonts}
\usepackage{algorithmic}
\usepackage{graphicx}
\usepackage{textcomp}
\usepackage{xcolor}
\usepackage{float}
\usepackage{algorithm2e}
\usepackage{subcaption}
\usepackage{svg}
\usepackage{fancyhdr}

\usepackage[style=ieee]{biblatex}
\newcommand{\thetitle}{Accelerating Triangle Counting with Real Processing-in-Memory Systems}
\title{\Large{\thetitle}}

\author{
{Lorenzo Asquini$^{1}$}\quad\hspace{0.5em}
{Manos Frouzakis$^{1}$}\quad\hspace{0.5em}
{Juan Gómez-Luna$^{2}$}\quad\hspace{0.5em}\\
{Mohammad Sadrosadati$^{1}$}\quad\hspace{0.5em}
{Onur Mutlu$^{1}$}\quad\hspace{0.5em}
{Francesco Silvestri$^{3}$}\vspace{0.7em}\\
{\small$^1$ETH Zürich}
\quad
{\small$^2$NVIDIA Research}
\quad
{\small$^3$University of Padova}
}

\begin{document}
\maketitle

\begin{abstract}

Triangle Counting (TC) is a procedure that involves enumerating the number of triangles within a graph. It has important applications in numerous fields, such as social or biological network analysis and network security. TC is a memory-bound workload that does not scale efficiently in conventional processor-centric systems due to several memory accesses across large memory regions and low data reuse. However, recent Processing-in-Memory (PIM) architectures present a promising solution to alleviate these bottlenecks.

Our work presents the first TC algorithm that leverages the capabilities of the UPMEM system, the first commercially available PIM architecture, while at the same time addressing its limitations. We use a vertex coloring technique to avoid expensive communication between PIM cores and employ reservoir sampling to address the limited amount of memory available in the PIM cores' DRAM banks. In addition, our work makes use of the Misra-Gries summary to speed up counting triangles on graphs with high-degree nodes and uniform sampling of the graph edges for quicker approximate results.
Our PIM implementation surpasses state-of-the-art CPU-based TC implementations when processing dynamic graphs in Coordinate List format, showcasing the effectiveness of the UPMEM architecture in addressing TC’s memory-bound challenges.

\end{abstract}
\section{Introduction}

The \emph{Triangle counting} (TC) problem consists in  counting all triangles (i.e., a clique of three nodes) in a given graph.
TC is a fundamental problem in graph processing, and it has many different applications, like spam detection \cite{spam_detection}, motif detection in biological networks \cite{motif_detection}, and sybil accounts detection \cite{sybil_account_detection}.
Graph-related problems, such as triangle counting, cannot scale efficiently in conventional processor-centric systems relying on CPUs or GPUs, where the limiting factor is the memory bandwidth. In graph analysis workloads, it is usually necessary to perform a large number of memory accesses across large memory regions, which leads to limited cache efficiency. Additionally, data reuse and computation per element of the graph are low, leading to the necessity of continuously fetching data from memory. 
This exacerbates the difference in time and energy needed for the computation and the data transfer and inhibits the seamless overlapping of memory transfers and computation. These aspects make it difficult to scale up this type of workload using conventional architectures, especially considering that simply increasing the core count for graph analysis proves ineffective due to the limited improvement in memory bandwidth \cite{ahn2015scalable}.
\par
Such problems could be alleviated using Processing-In-Memory (PIM) \cite{mutlu2023dac, mutlu2019processing, mutlu2025modern, ghose2019processing, mutlu2024memory, ghose2018enabling, mutlu2020intelligent, mutlu2019enabling, khan2024landscape, kautz1969cellular, stone1970logic}, a data-centric computing paradigm that places processing elements in close proximity to or within memory arrays, so that they can take advantage of higher memory bandwidth. In recent years, real-world PIM systems, like the one produced by UPMEM \cite{UPMEM_user_manual, gomez2022benchmarking, rhyner2024pimopt, giannoula2024accelerating, giannoula2024pygim, oliveira2022accelerating, alonsomarin2024bimsa, gogineni2024swiftrl, gupta2023evaluating, gómezluna2023experimental, item2023transpimlib, diab2023framework, giannoula2022sparsep, gomez2023evaluating, hyun2023pathfinding, hyun2024pathfinding, chen2023simplepim, gomez2021benchmarking, chen2023uppipe, lavenier2020variant, lavenier2016blast, jonatan2024scalability, nider2021case, khan2022cinm, lim2023design, bernhardt2023pimdb, baumstark2023adaptive, baumstark2023accelerating, das2022implementation, abumpimp2023, frouzakis2025pimdal}, have become commercially available, enabling new implementations for memory-bound applications.
The goal of this paper is to provide the first algorithm for triangle counting on real-world PIM architectures and to perform an extensive experimental evaluation. Specifically, we implement a triangle counting algorithm on the UPMEM PIM architecture.
\par
In order to make the best use of the UPMEM PIM architecture, it is necessary to address its limitations. One such limiting factor is the expensive cross-PIM core communications, which are routed through the host processor. To overcome this limitation and, at the same time, accelerate the execution by leveraging the thousands of available PIM cores, we employ a node coloring technique \cite{triangle_mapreduce} to split the problem into smaller sub-problems, allowing individual PIM cores to compute specific subsets of triangles in parallel without the need to communicate.
\par
Another constraint of the UPMEM PIM architecture that our TC implementation addresses is the limited size of the PIM cores' DRAM memory banks, which would not allow direct application of the algorithm to graphs with a large number of edges. For this reason, an approximate triangle counting technique \cite{TRIEST} is employed, leading to a randomized edge replacement within the available memory when there is not enough space for new ones. 
While this method might result in the loss of certain triangles during the counting phase, the final output undergoes statistical adjustments based on the total number of edges assigned to the PIM core, resulting in an approximate result with a low relative error.
\par
Transferring necessary edges to the PIM cores' DRAM banks requires the host processor to forward them from conventional DRAM. In order to lower the overhead caused by these transfers, our algorithm allows for uniform sampling of the graph's edges at the host level \cite{DUOLION_uniform_sampling}. This technique involves discarding some edges with a defined probability in order to lower the volume of data sent to the PIM cores. The resulting output when using this technique, after statistical adjustment, is an approximation of the number of triangles within the input graph.
\par
In triangle counting, employing an edge-iterator approach, such as the one used in our algorithm, can encounter slowdowns, particularly in the presence of nodes with a high degree. To address this issue, the Misra-Gries summary \cite{Misra-Gries} is used while reading the stream of edges of the input graph in order to approximately determine the highest degree nodes. This information is then used to iterate over lower-degree nodes first, optimizing the triangle counting process.

The contributions of this paper are:
\begin{itemize}
\item We introduce the first algorithm for Triangle Counting (TC) on the UPMEM PIM architecture. We propose different techniques to address the limitations of the current PIM architectures, such as the expensive cross-PIM core communications, the limited amount of memory available in the PIM cores' DRAM banks, and the need for data transfers between conventional DRAM and PIM cores through the host processor. Additionally, we address the limitations of the edge-iterator approach for triangle counting when dealing with graphs with high-degree nodes by using the Misra-Gries summary.
\item We perform an extensive experimental evaluation of our algorithm on a UPMEM PIM system under different settings, and show the trade-offs between performance and result quality.
We remark that the aim of this paper is not to outperform implementations on GPU and CPU for processing static graphs, but to explore the potential of a new hardware architecture from an algorithmic perspective and identify suitable applications.   
Nevertheless, we compare the \textit{exact} UPMEM implementation with state-of-the-art CPU and GPU implementations. While the CPU implementation, which makes use of the CSR matrix layout, and the GPU one outperform our solution, our implementation outperforms the state-of-the-art CPU implementation for triangle counting when processing dynamic input graphs in COO format.
\end{itemize}

\section{Background and Motivation}

In this section, we define the triangle counting (TC) problem and present the issues that arise from implementations of TC on processor-centric systems (using CPUs and GPUs) (Section \ref{ssec:triangle_counting_description}). We then provide an overview of PIM and the UPMEM PIM architecture (Section \ref{ssec:processing_in_memory_overview}).

\subsection{Triangle Counting}
\label{ssec:triangle_counting_description}

$G(V, E)$ is a simple, unweighted, and undirected graph where $V$ represents the set of vertices and $E$ represents the set of edges. Each vertex in the graph can be identified by a non-negative integer, and each edge in the graph can be identified by a pair of vertices $(u, v)$ such that $u,v \in V$. A triangle is defined as a set of three vertices ($u$, $v$, and $w$) such that any two of them are connected by an edge of the graph, meaning that the edges $(u, v)$, $(v, w)$, and $(w, u)$ are present in $E$. The triangle counting problem aims to determine the total number of triangles present in the graph $G$.
\par
Counting triangles encounters scalability issues in processor-centric systems due to limited DRAM memory bandwidth, which does not scale efficiently with an increase in processor cores. Both CPU (e.g., \cite{Map_JIK}) and GPU (e.g., \cite{TriCore}) implementations show suboptimal scaling when using multiple threads (CPU) or GPUs. Though the impact of limited scaling in memory bandwidth is less severe in GPUs, the inefficiency in scaling triangle counting persists as the number of CPU threads or GPUs increases.
\par
Algorithms for triangle counting have been extensively studied for several architectures, like standard CPU \cite{bader}, GPUs \cite{TriCore}, streaming \cite{TRIEST}, and parallel and distributed computing \cite{triangle_mapreduce,sanders}. 
Moreover, several algorithms have studied how to minimize the impact of memory accesses in the memory hierarchy \cite{PaghS14}. The most common technique for parallelizing and exploiting the memory hierarchy is based on vertex coloring. For reducing the computation time, a common heuristic technique leverages enumerating paths of length 2, in which the middle vertex has the lowest degree; the heuristic has been proven to have theoretical guarantees for power-law graphs \cite{berry}. Within approximation algorithms, sampling is among the most common techniques \cite{TRIEST}. In this paper, we build on some of these techniques to derive an efficient PIM solution.
\par
Previous works have explored triangle counting on PIM architectures. For example, \cite{other_TC_PIM_2020, other_TC_PIM_2022} proposes a TC accelerator utilizing Spin-Transfer Torque Magnetic RAM (STT-MRAM) arrays. Due to the theoretical nature of their architecture, the proposed accelerator is validated only through simulations. These works employ customized graph slicing and mapping techniques in order to compress the input graphs, as it is stated that popular graph representations, like COO, cannot be directly applied to in-memory computation. In contrast, our work presents a triangle counting algorithm implemented on a real PIM system that can be directly used with widely adopted graph formats, such as COO, without requiring novel compression methods, demonstrating the feasibility of using standard graph formats in PIM systems. 

\subsection{Processing-In-Memory (PIM)}
\label{ssec:processing_in_memory_overview}

The rise of Processing-in-Memory (PIM)  has introduced a paradigm shift in computing, placing processing units close to or within memory arrays, addressing the data movement bottleneck \cite{devaux2019true, chang2017understanding, seshadri2013rowclone, seshadri2015fast, chang2016low, seshadri2016buddy, seshadri2017ambit, seshadri2019dram, hajinazar2021simdram, seshadri2017simple, seshadri2016simple, cali2020genasm, hashemi2016accelerating, hashemi2016continuous, ahn2015scalable, ahn2015pim, boroumand2020practical, zhu2013accelerating, pugsley2014ndc, zhang2014top, farmahini2015nda, hsieh2016transparent, pattnaik2016scheduling, akin2015data, hsieh2016accelerating, lee2015bssync, gao2016hrl, chi2016prime, gu2016biscuit, kim2016neurocube, asghari2016chameleon, boroumand2016lazypim, liu2017concurrent, hassan2015near, nai2017graphpim, kim2017grim, fernandez2020natsa, singh2019napel, herruzo2021enabling, boroumand2021polynesia, giannoula2021syncron, besta2021sisa, asgari2021fafnir, denzler2023casper, oliveira2023dappa, sun2021abc, lee2022improving, dai2022dimmining, boroumand2019conda, drumond2017mondrian, dai2018graphh, zhuo2019graphq, oliveira2024mimdram, yuksel2024functionally, park2022flash, mansouri2022genstore, ghiasi2024megis, mao2022genpip, gao2019computedram, fernandez2024matsa, shahroodi2023swordfish, olgun2022pidram, ghiasi2022alp, oliveira2021damov, singh2020nero, boroumand2018google, boroumand2021google, oliveira2022heterogeneous, singh2021fpga, he2025papi, gu2025pim, ferreira2022pluto, gao2015practical, falahati2018origami, shelor2019reconfigurable, saikia2019k, kim2021gradpim, deng2018dracc, cho2020mcdram, shin2018mcdram, azarkhish2017neurostream, kwon2019tensordimm, ke2020recnmp, lee2021task, park2021high, kim2020mvid, wu2023pim, kang2023pim, liu2018processing, sun2020energy, imani2019floatpim, jiang2019cimat, schuiki2018scalable}. PIM's focus on proximity between processing elements and memory becomes even more important when also considering the widening gap between high computational performance and slow memory accesses.
PIM has been implemented in a multitude of architectures, like UPMEM \cite{gomez2022benchmarking}, SK Hynix AiM \cite{SK_Hynix_AiM}, Samsung HBM-PIM \cite{lee2021hardware, PIM_HBM2_samsung}, Samsung AxDIMM \cite{ke2021near}, and Alibaba HB-PNM \cite{niu2022184qps}.
\par
Our specific focus in this study is on the UPMEM PIM architecture \cite{UPMEM_user_manual, gomez2022benchmarking, rhyner2024pimopt, giannoula2024accelerating, giannoula2024pygim, oliveira2022accelerating, alonsomarin2024bimsa, gogineni2024swiftrl, gupta2023evaluating, gómezluna2023experimental, item2023transpimlib, diab2023framework, giannoula2022sparsep, gomez2023evaluating, hyun2023pathfinding, hyun2024pathfinding, chen2023simplepim, gomez2021benchmarking, chen2023uppipe, lavenier2020variant, lavenier2016blast, jonatan2024scalability, nider2021case, khan2022cinm, lim2023design, bernhardt2023pimdb, baumstark2023adaptive, baumstark2023accelerating, das2022implementation, abumpimp2023, frouzakis2025pimdal}, which is an example of a Processing-Near-Memory architecture, placing compute cores directly in the DRAM dies. An UPMEM DRAM DIMM consists of 16 DRAM banks on two ranks, each equipped with 8 processing elements called DPUs (as of March 2025).
\par
The DPUs are 32-bit in-order RISC-style, general-purpose processing cores. To reach optimal performance, they rely on deep pipelining and fine-grained multithreading capabilities, with software threads called tasklets; the architecture utilizes the Single Program Multiple Data (SPMD) programming model. Each DPU has access to its own 64-MB DRAM bank (MRAM), a 24-KB instruction memory (IRAM), and a 64-KB scratchpad memory (WRAM). The host CPU accesses the MRAM banks to allow transfers from the main memory to MRAM and vice versa. The architecture also has to rely on these transfers for inter-DPU communication since there is no direct channel between DPUs.
\par
Throughout our study, we adopted more generic terms applicable across various PIM systems and not exclusive to the UPMEM PIM architecture. For this reason, we use the terms \emph{PIM core, PIM thread, DRAM bank, scratchpad,} and \emph{CPU-PIM/PIM-CPU transfer}, which correspond to DPU, tasklet, MRAM bank, WRAM, and CPU-DPU/DPU-CPU transfer in UPMEM’s terminology \cite{UPMEM_user_manual}.
\section{Triangle Counting on a Real PIM Architecture}
In this section, we present our implementation of a triangle counting algorithm that makes use of the capabilities of a PIM architecture, detailing the strategies we use in order to overcome the limitations of such an architecture. In particular, we address the expensive cross-PIM core communications using a vertex coloring technique (Section \ref{ssec:edge_division_implementation}), the necessity of data transfer from conventional DRAM to PIM cores via the host processor by using uniform sampling at the host level (Section \ref{ssec:uniform_sampling_implementation}), and the limited amount of memory available in the PIM cores’ DRAM banks using reservoir sampling at the PIM core level (Section \ref{ssec:reservoir_sampling_implementation}). We then describe our optimized edge-iterator approach designed specifically for triangle counting within a PIM architecture (Section \ref{ssec:triangle_counting_implementation}) and our solution for handling high-degree nodes (Section \ref{ssec:misra_gries_implementation}).

\subsection{Edge Distribution Among PIM Cores}
\label{ssec:edge_division_implementation}

One of the limitations of different triangle counting algorithms on multi-processor systems is the need for a communication protocol between different cores to make sure that no triangle becomes uncounted or overcounted when distributing graph edges across multiple processing cores \cite{TC_distributed}. To avoid the complexities and cost of inter-PIM core communication, a coloring-based partitioning of the edges is performed to divide the workload across the PIM cores.
\par
We color each node uniformly at random using $C \geq 1$ colors; colors are identified by a number from 0 to $C-1$. Then each PIM core is assigned a distinct triplet of ordered colors, which describes one of the potential color configurations of a triangle inside the graph.
\par
When an edge is read from the input file, its nodes are colored through a hash function that, given $C$ colors, allows for an even distribution of colors across the graph's nodes.
We used $h_C(u) = ((a \cdot u + b) \mod p) \mod C$, where $p$ is a suitably large prime number, $a$ is a random integer in $[1, p-1]$, and $b$ is a random integer in $[0, p-1]$.
\par
\begin{figure}
    \centering
    \includegraphics[width=0.45\textwidth]{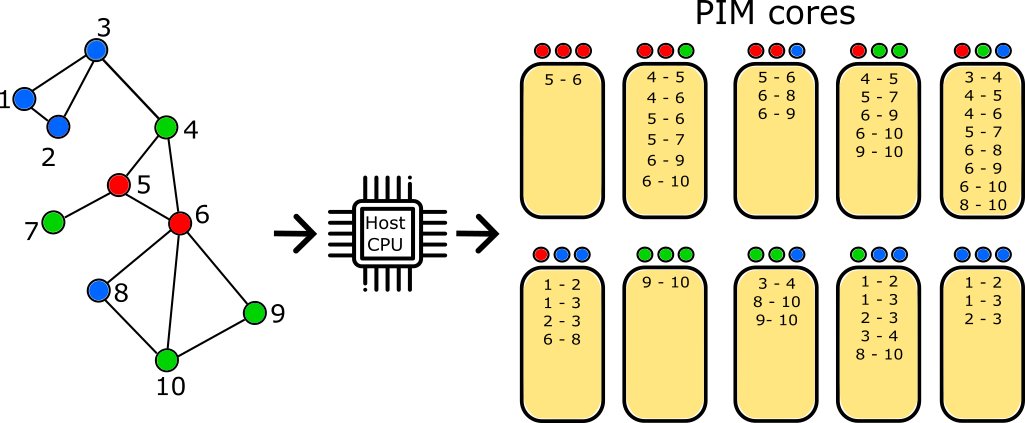}
    \caption{Partitioning of the graph's edges among the PIM cores through the host processor.}
    \label{fig:edge_division}
\end{figure}

After coloring  nodes connected by an edge, the edge is assigned to one or more compatible PIM cores, determined by the colors present within their corresponding triplets. The edge's compatibility with a triplet and its corresponding PIM core relies on the presence of a match between the colors in the edge and the colors in the triplet. For example, the triplet with colors $(0, 1, 2)$ has the following compatible pairs of colors: $(0, 1)$, $(1, 2)$, and $(0, 2)$.
Figure \ref{fig:edge_division} exemplifies the partitioning of edges among the PIM cores. After the nodes have been colored, they are transferred by the host processor to the PIM cores according to the triplets of colors of the destination PIM cores. 

We now analyze the drawbacks of this technique.

\paragraph{Redundant counting} Triangles formed by nodes all with the same color might be counted by multiple PIM cores because subsets of edges assigned to different PIM cores might share the edges that are needed to identify the same triangles. For example, a triangle whose nodes have the colors $(0, 0, 0)$ is counted both by the PIM core which is assigned the triplet $(0, 0, 0)$ and by the PIM core which is assigned the triplet $(0, 0, 1)$.
This issue can be addressed by knowing that the triangles counted multiple times are counted exactly by $C$ different PIM cores. When a specific triplet assigned to a PIM core contains only one color, that core becomes responsible for counting the triangles with nodes colored only with that color. This count can be used to correct the final result, knowing that other $C-1$ PIM cores will also have counted these same triangles.

\paragraph{Edge Duplication} Each edge is duplicated $C$ times, where $C$ denotes the number of colors used. Although edge duplication limits the size of the graphs that could be processed, we implemented a reservoir sampling technique (described in Section \ref{ssec:reservoir_sampling_implementation}) that allows for the analysis of graphs with any number of edges by approximating the final result.

\paragraph{Uneven Edge Distribution} Different PIM cores are assigned a different number of edges. Defining $N$ as the number of edges assigned to the PIM cores that handle triplets with a single color, the PIM cores handling triplets with two different colors will need to handle $3 \cdot N$ edges, and the ones handling the triplets with three different colors will need to handle $6 \cdot N$ edges. This is because, for triplets with a single color, each edge has only one possible color assignment. For triplets with two colors, $N$ edges are fixed in their color arrangement, while another $N$ edges can appear in either of two possible permutations. Similarly, for triplets with three colors, each combination of two colors within the triplet has two possible permutations, resulting in a total of $6N$ edges. In particular, $C$ PIM cores are assigned $N$ edges, $2 \cdot \binom{C}{2}$ PIM cores $3 \cdot N$ edges, and $\binom{C}{3}$ PIM cores $6 \cdot N$ edges. This means that with increasing $C$, due to the nature of the binomial coefficient, a majority of the cores will be processing $6 \cdot N$ edges, keeping the load balancing even.
\par
In our implementation, the host CPU reads a file containing the COO representation of the graph. This representation illustrates the unweighted graph matrix using a list of (row, column) tuples, where each tuple denotes the source and target nodes for an edge within the graph.
\par
Each host CPU thread manages an array of edges per PIM core, which are populated according to the specific triplet assigned to each PIM core. Once all edges have been processed, each thread transfers its different batches of edges to all PIM cores in parallel. When a PIM core receives the edges, it copies them to the correct location in the DRAM bank or applies reservoir sampling \ref{ssec:reservoir_sampling_implementation} if space is insufficient.

\subsection{Approximate Triangle Counting Through Uniform Sampling at the Host Level}
\label{ssec:uniform_sampling_implementation}

To reduce the volume of data to process and the execution time, we adopt a strategy of uniform edge sampling at the host level, although at the expense of generating approximate results. This method involves discarding an edge with a probability of $1-p$ as we read the file from the input file.
\par
Once the host retrieves the resulting number of triangles from the PIM cores, composed of the edges actually considered (that may be all of the edges of the graph if $p=1$), the result is corrected to account for the discarded edges. If $p$ represents the probability of an edge being considered, the probability that all three edges of a triangle are included is $p^3$. 
Therefore, an unbiased estimator is given by dividing the number of counted triangles by $p^3$ (see e.g., \cite{DUOLION_uniform_sampling}).
\par
This method accelerates both host code and kernel execution on the PIM cores. The process of discarding edges during graph reading leads to fewer edges being inserted into the batches destined for the PIM cores. As a result, the creation of the batches and their transfer time to the PIM cores decreased. Furthermore, the reduction in edges processed by each PIM core translates to decreased time required for triangle counting.
\par
Note that this technique can be applied concurrently with Reservoir Sampling (Subsection \ref{ssec:reservoir_sampling_implementation}).

\subsection{Approximate Triangle Counting Through Reservoir Sampling at the PIM Cores Level}
\label{ssec:reservoir_sampling_implementation}

To address potential PIM DRAM banks capacity limitations when dealing with graphs with a large number of edges, a technique called reservoir sampling is employed when the allocated memory in a PIM core's DRAM is insufficient to accommodate all edges. This method allows for approximating the total triangle count while working within a specified memory constraint \cite{TRIEST}.
\par
Let $M$ be the maximum amount of edges that a PIM core can store in a sample of edges $S$ inside its DRAM bank. Considering the $t$-th edge $e_t = (u, v)$ received by a DPU:

\begin{itemize}
    \item If $t \leq M$, then the edge $e_t$ is deterministically inserted in $S$.
    \item If $t > M$, a biased coin with a heads probability of $M/t$ is tossed. If the outcome is heads, an edge $(w, z)$ is chosen uniformly at random from $S$, and it is replaced by the newly encountered edge $(u, v)$. Otherwise, $S$ is not modified. The COO representation of the sub-graph assigned to a the PIM core in its DRAM bank facilitates the edge replacement.
\end{itemize}

This might lead to the loss of some triangles. To obtain an approximate result, the value $T$ of triangles counted by each PIM core is adjusted by dividing it by a factor $p = \frac{M \cdot (M-1) \cdot (M-2)}{t \cdot (t-1) \cdot (t-2)}$, where $M$ represents the maximum number of edges that fit within the edge sample and $t$ denotes the total number of edges that were assigned to the particular PIM core.
\par
Note that this technique can be applied concurrently with Uniform Sampling (Subsection \ref{ssec:uniform_sampling_implementation}).

\subsection{Triangle Counting}
\label{ssec:triangle_counting_implementation}

Upon receiving the necessary edges from the host processor, each PIM core will possess a sample $S$ of edges within its DRAM bank. As the first step of the triangle counting phase, the edges are ordered in ascending order using the IDs of their nodes, ensuring that for every edge $(u,v)$, the condition $u < v$ holds, which is required for the technique to function correctly. The following comparison is used:

\setlength{\abovedisplayskip}{0pt}
\begin{equation*}
    (u, v) < (w, z) \leftrightarrow u < w \vee (u = w \wedge v < z)
\end{equation*}
\setlength{\abovedisplayskip}{\baselineskip}

\begin{figure}
    \centering
    \includegraphics[width=0.45\textwidth]{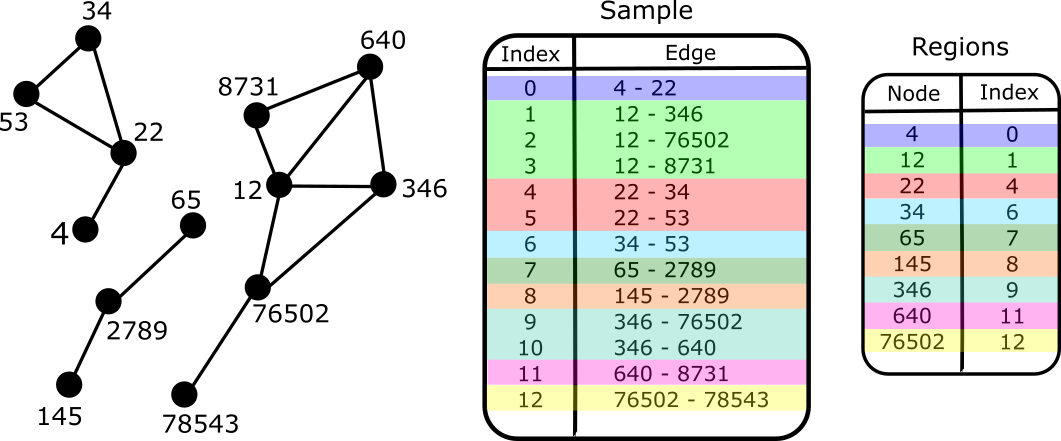}
    \caption{Representation of a subgraph assigned to a PIM core inside its DRAM bank.}
    \label{fig:regions}
\end{figure}

After ordering the edges, it is possible to identify regions of edges inside the sample $S$ that share a common first node, with the second node being ordered in ascending order and representing one of the neighbors of the first node. Within the DRAM bank of each PIM core, for every such region, an entry is generated. This entry contains the identifier of the first node of the edges within the region and the index denoting the first edge of this particular region within the sample $S$.
Figure \ref{fig:regions} illustrates the representation of a subgraph that may be assigned to a DPU using a COO representation within the sample $S$, accompanied by the corresponding indexing table delineating its edge regions.

Following these preparatory steps, it is possible to start counting the triangles. Each PIM thread retrieves a buffer of edges from the sample $S$ into the scratchpad memory until there are no more edges to consider. For each edge $(u, v)$ within this buffer, the thread performs a binary search in the DRAM to locate edges originating from node $v$ using the indexed regions described previously. If no such edge is found, meaning there is no ordered edge where $v$ is the first node, the thread proceeds to the next edge in the local buffer. However, if edges starting with node $v$ are located, they are loaded into the scratchpad memory.
\par
Our implementation of triangle counting employs a merge-like approach, making use of the node order within each edge. Given two edges $(u, w)$ and $(v, z)$, a comparison is made between nodes $w$ and $z$. If $w = z$, it signifies that nodes $u$ and $v$ share a common neighbor, implying the existence of a triangle $(u, v, w = z)$ to be counted. In this case, both the next edge with $u$ as the first node and the next edge with $v$ as the first node are considered. If $w < z$, the next edge with $u$ as the first node is considered. If $w > z$, the next edge with $v$ as the first node is considered. When all edges starting with $u$ or $v$ loaded in the scratchpad memory have been processed, new edges are loaded from the DRAM bank. 
\par 
After each PIM thread has concluded the counting, all the returned values are summed up for the final result, $T$. It may be necessary to correct the result in the event that some edges were replaced, as described in Section \ref{ssec:reservoir_sampling_implementation}.
For the final triangle count, the host processor gathers all the partial counts from the PIM cores and adds them up.

\subsection{Handling High Degree Vertices}
\label{ssec:misra_gries_implementation}

The technique for triangle counting described previously operates by processing edges based on the ID of their first node, starting with edges with lower first node IDs. For an edge $(u, v)$ where node $u$ has a high degree, the number of neighbors $v$ to consider can be quite large. The larger the number of neighbors, the more time it takes to process all edges originating from $u$, primarily due to the substantial number of comparisons needed to identify matching neighbors of both $u$ and $v$. Depending on the maximum node degree within a graph, this may significantly reduce performance, even when the number of edges and triangles is moderate.
\par
\begin{figure}
    \centering
    \includegraphics[width=0.45\textwidth]{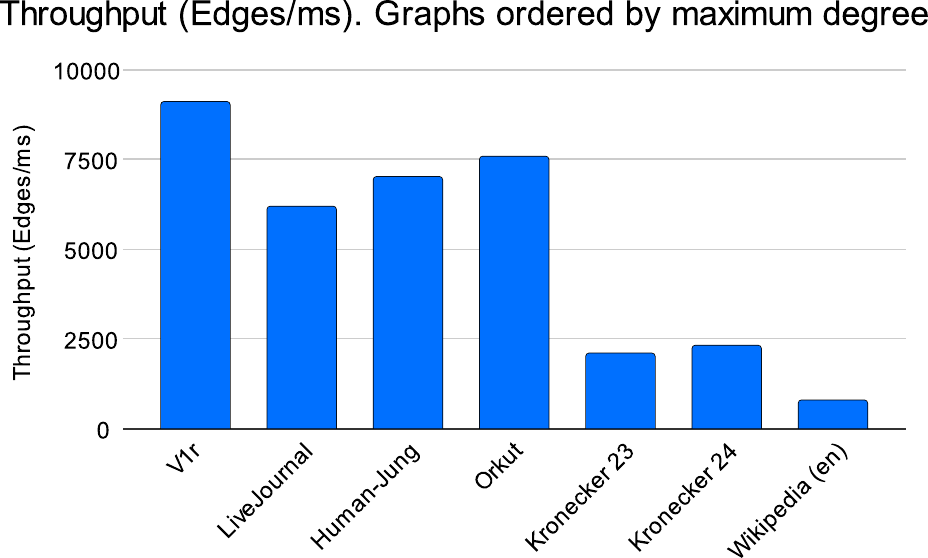}
    \caption{Throughput (edges per millisecond) when counting triangles in different graphs, ordered by their maximum node's degree (lowest first).}
    \label{fig:throughput}
\end{figure}

Figure \ref{fig:throughput} shows the throughput measured in edges per millisecond while employing our PIM implementation for triangle counting across various graphs. Considering that the ordering of the graphs in the plot is based on the maximum degree observed within their nodes, it is possible to notice a considerable difference in throughput between the first four graphs, which have a maximum degree in the tens of thousands, and the last three graphs, which have a maximum degree in the hundreds of thousands or millions. This illustrates the considerable impact that high-degree nodes can have on performance when counting the number of triangles in a graph using our proposed method.

To improve performance in graphs with high-degree nodes, our proposal utilizes the Misra-Gries summary algorithm \cite{Misra-Gries}, a common data stream algorithm to extract frequent items. This approach approximately identifies high-degree nodes, allowing for the reordering of only the edges associated with these nodes. When this technique is used, each host thread, while processing its assigned section of the input graph, employs the Misra-Gries summary algorithm to identify the most frequent nodes.
\par
The frequencies using the Misra-Gries summary are determined as follows:
\begin{itemize}
    \item If there is an entry with the current node in the hash table, its frequency is increased by one.
    \item If there is not an entry with the current node in the hash table and the total number of entries is less than the parameter $K$, a new entry is created for that node with the frequency set to 1.
    \item If there is no entry with the current node in the hash table and the total number of entries is equal to the parameter $K$, all the frequencies of the entries in the hash table are decreased by one. Entries with a frequency of zero are removed. This ensures that there are no more than $K$ entries in the hash table at any given moment.
\end{itemize}

For a host thread handling a section of the input graphs with $n$ edges, each node that has a degree higher than $n/K$ in that section is guaranteed to be present in the hash table at the end of the edge stream for that thread.
\par
Another critical parameter is denoted as $t$, representing the numbers of top-degree nodes to be remapped into new IDs within the PIM cores and that are therefore sent to them. In the PIM cores, before the triangle counting phase and specifically before the ordering step, edges associated with high-degree nodes are mapped to new IDs. Nodes with the highest degrees are remapped to new IDs outside the initial set of IDs. In particular, the most frequent node is assigned to the highest new ID. This ensures that the neighbors of the most frequent nodes that need to be traversed to find triangles are low or zero in the case of the most frequent node. This lowers the amount of matchings to be checked while guaranteeing that all triangles are counted, improving the triangle counting performance.
\section{Evaluation}

This section presents the evaluation of our triangle counting algorithm designed for a real-world PIM system, comparing it with implementations on CPU and GPU that accept a COO-formatted input graph, with a particular focus on dynamic graphs. Following a description of the evaluation methodology (Section \ref{ssec:evalutation_methodology}), we evaluate the performance of the PIM implementation while varying the number of colors and thus the number of PIM cores employed (Section \ref{ssec:pim_core_scaling_evaluation}). Afterwards, the performance improvements that come from using the Misra-Gries summary are shown (Section \ref{ssec:misra_gries_evaluation}), as well as the benefits that can come from uniform sampling (Section \ref{ssec:uniform_sampling_evaluation}) and the performance of the proposed algorithm when reservoir sampling is used (Section \ref{ssec:reservoir_sampling_evalutation}). At last, our implementation of the TC algorithm on a PIM architecture is compared against state-of-the-art CPU and GPU implementations capable of working with COO-formatted input graphs, both considering static and dynamic graphs (Section \ref{ssec:CPU_GPU_comparison}).

\subsection{Methodology}
\label{ssec:evalutation_methodology}

Table \ref{tab:graphs_info} presents the graphs in COO representation that were used in the evaluation process. The graphs were preprocessed by: removing duplicate edges and self-loops (i.e. each edge $(u, v)$ where $u = v$);
shuffling the resulting graph using the command line utility \verb|shuf|.

\begin{table}
\scriptsize
\centering
\begin{tabular}{|c|c|c|c|}
\hline
\textbf{Graph} & \textbf{$|E|$} & \textbf{$|V|$} & \textbf{Triangles} \\ \hline
Kronecker 23 \cite{graph500}  & 129,335,985 & 4,609,311 & 4,675,811,428 \\ \hline
Kronecker 24 \cite{graph500}  & 260,383,358 & 8,870,393 & 10,285,674,980  \\ \hline
V1r \cite{suite_sparse}       & 232,705,452 & 214,005,017  & 49  \\ \hline
LiveJournal \cite{SNAP}       & 42,851,237  & 4,847,571   & 285,730,264   \\ \hline
Orkut \cite{SNAP}             & 117,185,083 & 3,072,441  & 627,584,181   \\ \hline
Human-Jung \cite{network_repository}  & 267,844,669  & 784,262  & 41,727,013,307  \\ \hline
WikipediaEdit \cite{konect}   & 255,688,945 & 42,541,517 & 881,439,081    \\ \hline
\end{tabular}
\caption{Graphs used in the evaluations.}
\label{tab:graphs_info}
\end{table}

The evaluation system for the PIM implementation and the CPU implementation features:
\begin{itemize}
    \item 20 PIM-enabled DIMMs (codename P21), providing 2560 PIM cores (DPUs).
    \item Four DRAM memory DIMMs, resulting in a total memory of 256GB.
    \item Two Intel Xeon Silver 4215 CPUs \cite{xeon_silver_4215} in a dual socket configuration.
\end{itemize}

The system for evaluating the GPU implementation features:
\begin{itemize}
    \item Two Intel Xeon Gold 5118 CPUs \cite{xeon_gold_5118} in a dual socket configuration.
    \item 192GB of DRAM memory.
    \item One Nvidia A100 \cite{nvidia_a100}, with 80GB of VRAM.
\end{itemize}

The times in the subsequent plots show the average time derived from five measurement runs. Error bars are not included in the plots because the coefficient of variance is consistently below 5\%, and most of the time below 2\%.
\par
The time for each run using the UPMEM PIM system is divided into three sections:
\begin{itemize}
    \item \textit{Setup Time:} Preparation of the PIM cores and the host processor, involving PIM cores allocation, kernel loading, and the variable initialization for each PIM core. This includes loading the graph file into host memory and allocating arrays for the batch creation.
    \item \textit{Sample Creation Time:} Reading the input graph and creating the batches for each PIM core, with their subsequent transfers. A sample $S$ is therefore created in the DRAM bank of each PIM core, with the use of reservoir sampling if necessary.
    \item \textit{Triangle Count Time:} Organizing the samples in the PIM cores' DRAM bank and executing triangle counting. Simultaneously, the host processor frees previously allocated memory. This phase concludes when the results have been gathered by the host processor, the final result has been returned, and the PIM cores have been freed.
\end{itemize}

When evaluating the PIM implementation, the host processor uses 32 threads, while each PIM core uses 16 PIM threads.
\par
The source code is available at \url{https://github.com/CMU-SAFARI/PIM-TC}.

\subsection{PIM Cores Scaling}
\label{ssec:pim_core_scaling_evaluation}

The performance of the algorithm we present depends on the number of colors denoted as $C$, which determines the number of PIM cores utilized, which is equal to $\binom{C+2}{3}$.
\par
During this evaluation, the baseline number of PIM cores used varies depending on the input graph. This is due to the fact that, for bigger graphs, using fewer PIM cores leads to approximate results because of the use of reservoir sampling. With the goal of analyzing the scalability when changing the number of PIM cores while calculating the exact number of triangles, it was necessary to test different numbers of PIM cores for each graph.
\par
\begin{figure*}
    \centering
    \begin{subfigure}[h]{0.06\linewidth}
        \centering
        \includegraphics[width=\linewidth]{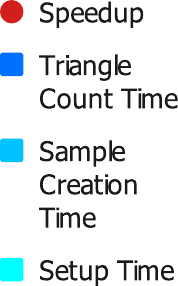}
    \end{subfigure}
        \hfill
    \begin{subfigure}[h]{0.9\linewidth}
        \centering
        \includegraphics[width=\linewidth]{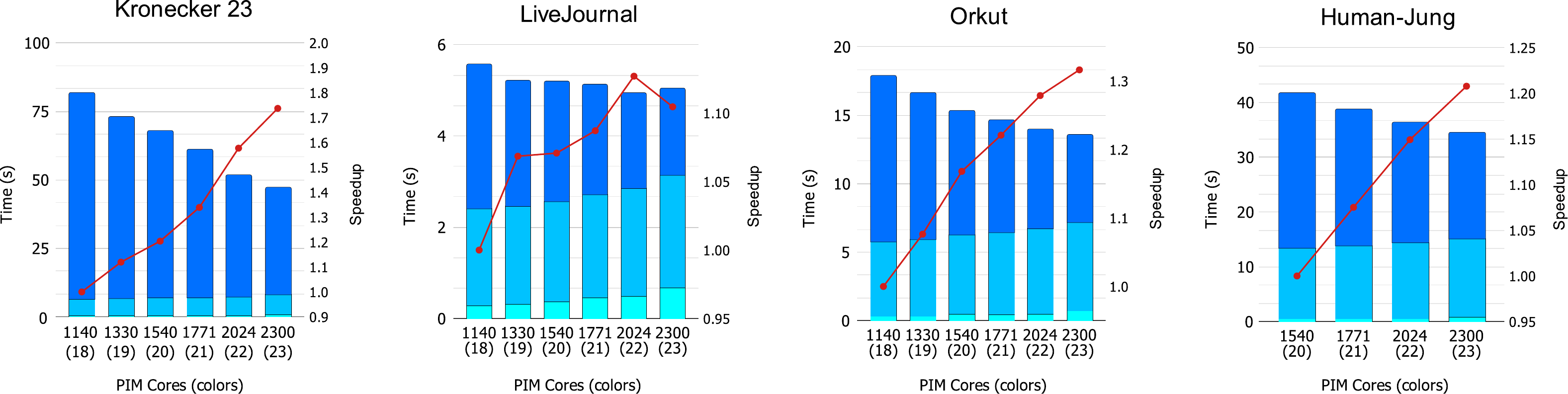}
    \end{subfigure}

    \caption{Performance scaling when changing the number of PIM cores and the number of colors (between brackets) used to partition the graphs. The number of threads used by the host processor is 32.}
    \label{fig:pim_core_scaling}
\end{figure*}

Figure \ref{fig:pim_core_scaling} shows the time required and the speedup compared to the lowest number of PIM cores tested for each specific graph when computing the exact count of triangles in four different graphs for varying colors and PIM core counts. In most scenarios, the execution time for triangle counting in different graphs decreases with an increasing number of PIM cores. However, when analyzing the \emph{LiveJournal} graph, the algorithm performs faster with fewer PIM cores than the maximum available. This is due to the fact that the number of edges in the graph is low, and the improvements from using more PIM cores are offset by the associated overhead, consisting of more time spent on PIM core allocation and data transfers to and from the PIM cores.
\par
For the following evaluations, the setup time will not be considered, and therefore the configuration that uses the highest valid number of DPUs in the system will be used (23 colors, 2300 DPUs).

\subsection{Misra-Gries Summary}
\label{ssec:misra_gries_evaluation}

In order to speed up the execution time on graphs with high-degree nodes, the Misra-Gries summary is used while varying two key parameters, $K$ and $t$. The former affects the search accuracy for the most frequent nodes, with a higher value resulting in a more accurate estimation of the heavy-hitter nodes. The latter directly affects the performance in the triangle counting phase, determining the number of nodes that are remapped in the PIM cores.
\par
Table \ref{tab:node_degrees} shows the maximum node degree, the average node degree and the global clustering coefficient in the graphs in our evaluations. It is possible to notice that three graphs (\textit{Kronecker 23}, \textit{Kronecker 24}, and \textit{WikipediaEdit}) have a maximum node degree at least an order of magnitude higher than the other graphs.

\begin{table}
\centering
\footnotesize
\begin{tabular}{|c|c|c|c|}
\hline
\textbf{Graph} & \textbf{Max degree} & \textbf{Average degree} & \textbf{Global Clustering} \\ 
& & & \textbf{Coefficient} \\ \hline
Kronecker 23   & $257,484$                      & $56.12$                & $0.0209$                \\ \hline
Kronecker 24   & $407,017$                      & $58.71 $               & $0.0173$                \\ \hline
V1r            & $8$                            & $2.17$                 & $4.784 \cdot 10^{-7}$   \\ \hline
LiveJournal    & $20,333$                       & $17.68$                & $0.1179$                \\ \hline
Orkut          & $33,313$                       & $76.28$                & $0.0413$                \\ \hline
Human-Jung     & $21,743$                       & $683.05$               & $0.2944$                \\ \hline
WikipediaEdit  & $3,026,864$                    & $12.02$                & $7.827 \cdot 10^{-5}$   \\ \hline
\end{tabular}
\caption{Maximum node degree, average node degree and global clustering coefficient in the considered graphs.}
\label{tab:node_degrees}
\end{table}

\begin{figure*}
    \centering
    \begin{subfigure}[h]{0.1\linewidth}
        \centering
        \includegraphics[width=\linewidth]{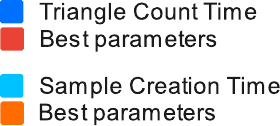}
    \end{subfigure}
    \hfill
    \begin{subfigure}[h]{0.875\linewidth}
        \centering
        \includegraphics[width=\linewidth]{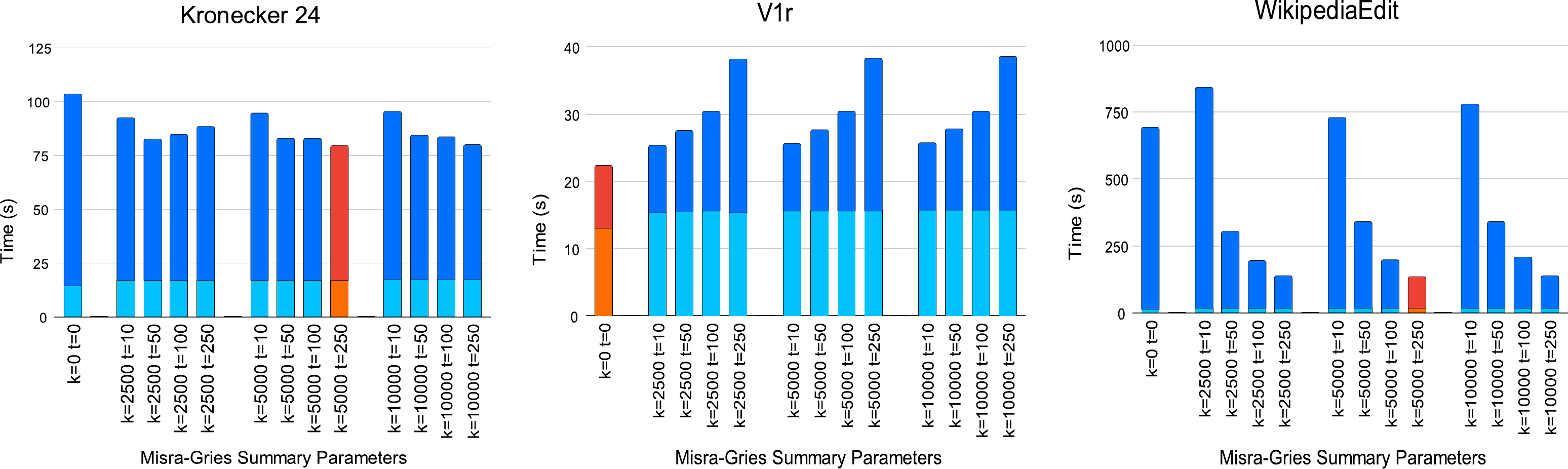}
    \end{subfigure}

    \caption{Performance results when changing the parameters $K$ and $t$ used by the Misra-Gries summary.}
    \label{fig:misraGries_plots}
\end{figure*}

Figure \ref{fig:misraGries_plots} illustrates the time needed to compute the exact count of triangles in different graphs while varying the values of $K$ and $t$ used by the Misra-Gries summary. We notice that this technique has no advantages on graphs with lower-degree nodes and, on the contrary, increases the execution time. In particular, the most expensive operation is the remapping of the most frequent nodes, as can be seen by the increase in computation time when increasing the value of $t$.
\par
On the other hand, graphs with higher degree nodes greatly benefit from using the Misra-Gries summary, both by increasing the accuracy for higher values of $K$, and by increasing the effect of the remapping on the sample in the PIM cores with higher values for the parameter $t$. Depending on the maximum frequency of the nodes and the number of heavy hitters in the graph, after a certain point, however, there may be no additional benefits from increasing the values $K$ and $t$, with diminishing performance improvements that are hidden by the additional computation cost.
\par
The best performing parameters for the Misra-Gries summary for each graph will be used in the following evaluations.

\subsection{Uniform Sampling}
\label{ssec:uniform_sampling_evaluation}

In order to find the benefits and downsides of uniform sampling, we ran our triangle counting algorithm on different graphs, keeping different percentages $p$ of the edges of the graphs, with $p \in [0.5, 0.25, 0.1, 0.01]$.
\par
Table \ref{tab:uniform_sampling_relative_error} shows the relative error observed in the counted triangles while varying the percentage of edges considered from the input graphs. Even when considering only $1 \%$ of the total edges of the input graphs, the relative error typically remains below $2.5\%$ in most instances. However, \textit{V1r} is an exception, with its edges forming only $49$ triangles. Consequently, even a minimal elimination of edges removes a substantial portion of triangles relative to the small total. This disparity can consequently lead to considerably higher errors, particularly when no triangles are counted (that leads to a relative error of 100\%). 
\par
Using this technique can lead to a speedup of up to 80 times when $p=0.01$. These significant speedups are achievable in graphs with a substantial number of edges, where the overhead of creating and transferring batches of edges to the PIM cores in the host processor has a minimal impact on the total execution time. 

\begin{table}
\centering
\footnotesize
\begin{tabular}{|p{1.8cm}|p{1.15cm}|p{1.15cm}|p{1.15cm}|p{1.1cm}|}
\hline
\textbf{Graph} & \textbf{0.5} & \textbf{0.25} & \textbf{ 0.1} & \textbf{0.01} \\ \hline
Kronecker 23    & 0.025\%  & 0.050\%   & 0.128\%  & 0.133\%    \\ \hline
Kronecker 24    & 0.009\%  & 0.096\%   & 0.003\%  & 0.379\%    \\ \hline
V1r             & 2.857\%  & 21.633\%  & 100\%    & 100\%      \\ \hline
Orkut           & 0.007\%  & 0.014\%   & 0.017\%  & 2.050\%    \\ \hline
LiveJournal     & 0.093\%  & 0.031\%   & 0.087\%  & 2.379\%    \\ \hline
Human-Jung      & 0.010\%  & 0.012\%   & 0.024\%  & 0.397\%    \\ \hline
WikipediaEdit   & 0.044\%  & 0.430\%   & 0.810\%  & 1.614\%    \\ \hline
\end{tabular}
\caption{Relative error in the counted triangles when varying the percentage $p$ of edges of the input graphs considered, for $p\in [0.5, 0.25, 0.1, 0.01]$.}
\label{tab:uniform_sampling_relative_error}
\end{table}

\subsection{Reservoir Sampling}
\label{ssec:reservoir_sampling_evalutation}

Considering $|E|$ as the number of edges in a graph and $C$ as the number of colors utilized, the maximum \emph{expected} number of edges assigned to a PIM core can be calculated as $\frac{6}{C^2} \cdot |E|$. To simulate scenarios where reservoir sampling might be necessary, the sample size within the PIM cores was limited to a percentage $p$ of the expected necessary size, with $p \in [0.5, 0.25, 0.1, 0.01]$.
\par
Table \ref{tab:reservoir_sampling_relative_error} showcases the observed relative error in counted triangles while modifying the sample size in the PIM cores based on the maximum anticipated number of edges allocated to the PIM cores across various graphs. The relative error remains consistently low, staying below $0.6\%$ across most cases examined, except for the graph \textit{V1r}, which experiences a similar issue as previously described due to its limited number of triangles.

\begin{table}
\centering
\scriptsize
\begin{tabular}{|p{1.65cm}|p{0.85cm}|p{0.85cm}|p{0.85cm}|p{0.85cm}|}
\hline
\textbf{Graph} &  \textbf{0.5} & \textbf{0.25} & \textbf{0.1} & \textbf{0.01} \\ \hline
Kronecker 23      & 0.003\%  & 0.002\%   & 0.014\%  & 0.514\%   \\ \hline
Kronecker 24      & 0.010\%  & 0.010\%   & 0.011\%  & 0.681\%   \\ \hline
V1r               & 24.49\%  & 15.51\%   & 309.4\%  &  100\%    \\ \hline
Orkut             & 0.007\%  & 0.013\%   & 0.022\%  & 0.004\%   \\ \hline
LiveJournal       & 0.023\%  & 0.056\%   & 0.207\%  & 0.283\%   \\ \hline
Human-Jung        & 0\%      & 0\%       & 0.001\%  & 0.594\%   \\ \hline
WikipediaEdit     & 0.052\%  & 0.283\%   & 0.254\%  & 1.030\%   \\ \hline
\end{tabular}
\caption{Relative error in the counting triangles when varying the size of the samples $S$ in the PIM cores.  The sample size within the PIM cores was limited to a percentage $p$ of the expected necessary size, with $p \in [0.5, 0.25, 0.1, 0.01]$.}
\label{tab:reservoir_sampling_relative_error}
\end{table}

In our tests using reservoir sampling, the time for triangle counting  decreases due to the lower number of edges to consider when counting triangles on each PIM core, while the time needed to create the samples inside the PIM cores increases due to the need for edge replacements in the samples $S$ once they are full. Our tests show that the time required for both these two phases between $p = 0.5$ and $p = 0.01$ can increase when the sample creation phase is a significant portion of the total execution time. 
On the other hand, when the triangle counting phase dominates the execution time, as in graphs like \texttt{Kronecker 23}, \texttt{Kronecker 24}, and \texttt{WikipediaEdit}, the total execution time tends to decrease.
\par
While uniform sampling could have offered similar results, even with a significant performance improvement, reservoir sampling is advantageous since it generally yields a lower final result error, and it is adaptive and based on the size of the PIM DRAM banks (i.e., not requiring any manual change in the percentage $p$ of edges to consider).

\subsection{Comparison to CPU and GPU Implementations}
\label{ssec:CPU_GPU_comparison}

To compare our triangle counting implementation on the UPMEM PIM system against state-of-the-art TC implementations on traditional processor-centric systems, we selected the best-performing CPU implementation \cite{Map_JIK, CPU_TC_implementation} and GPU implementation \cite{cugraph} that directly accept, without requiring any extra modifications, COO-formatted input graphs, a format widely used across many publicly available graph datasets, in order to have a common input format for all implementations. All the input graphs have been preprocessed as described in Section \ref{ssec:evalutation_methodology}, and the time for this operation is not considered in the comparison times.
\par
\begin{figure}
    \centering
    \includegraphics[width=0.9\linewidth]{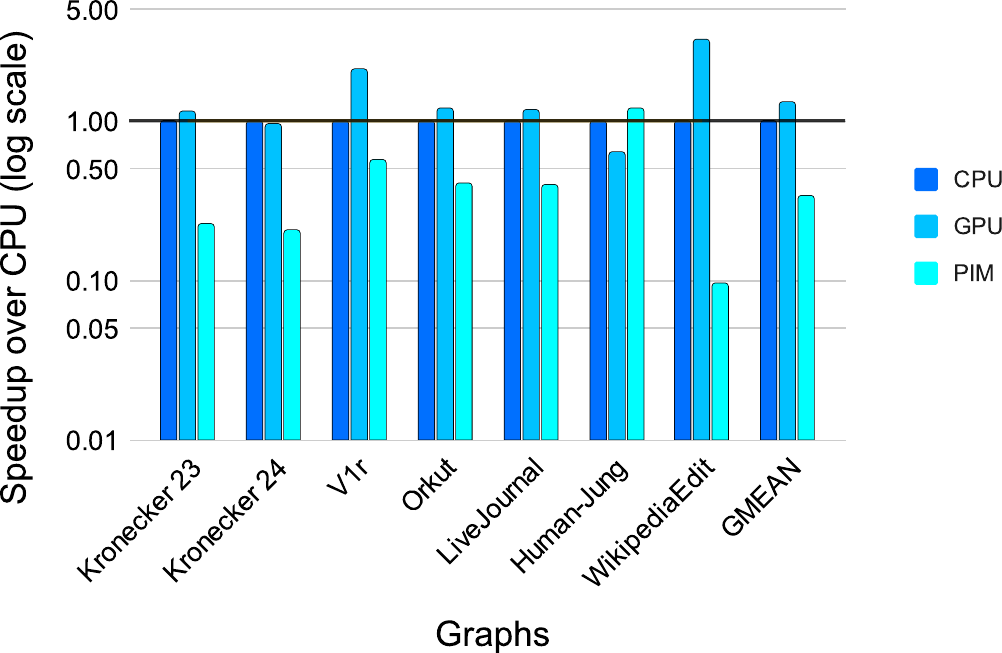}
    \caption{Speedup of PIM and GPU implementations over the baseline CPU implementation when counting the exact number of triangles given different static input graphs.}
\label{fig:speedupStatic}
\end{figure}

Figure \ref{fig:speedupStatic} illustrates the speedup achieved by both the PIM and GPU implementations in comparison to the CPU implementation, which serves as the baseline, when considering only the time required in the different platforms to count the exact number of triangles once the graph was in memory. While the CPU implementation can accept graphs in COO format, it necessitates an internal conversion to CSR format, though the conversion time is excluded from this comparison.
\par
The GPU implementation consistently outperforms both the CPU and PIM implementations, with the CPU typically following in terms of performance. The PIM implementation generally lags behind, except in the case of the \texttt{Human-Jung} graph, where it surpasses the other methods due to the graph's high triangle count and low maximum node degree. The CPU implementation benefits from using the CSR format internally, and considering that the initial conversion time from COO to CSR is not included in this comparison, it provides a performance edge in the triangle counting task.

Graphs in COO format are widely used in triangle counting applications, particularly for dynamic graphs, because they allow for easy updates by simply appending new edges to the existing list \cite{dynamic_triangle_counting}, unlike other formats such as CSR, which require a more complex update process. To demonstrate the applicability of the PIM implementation in a dynamic graph environment, we chose to simulate a workload using the \texttt{WikipediaEdit} graph, which showed the worst performance for the PIM implementation compared to the CPU implementation, as an example of the worst-case scenario for the PIM implementation.
\\
In our simulation, we divided the initial graph into 10 smaller subgraphs, aiming to count triangles iteratively. Each step involved merging a new subgraph with the existing graph, effectively simulating 10 dynamic updates and corresponding counting phases.
\par
\begin{figure}
    \centering
    \includegraphics[width=0.9\linewidth]{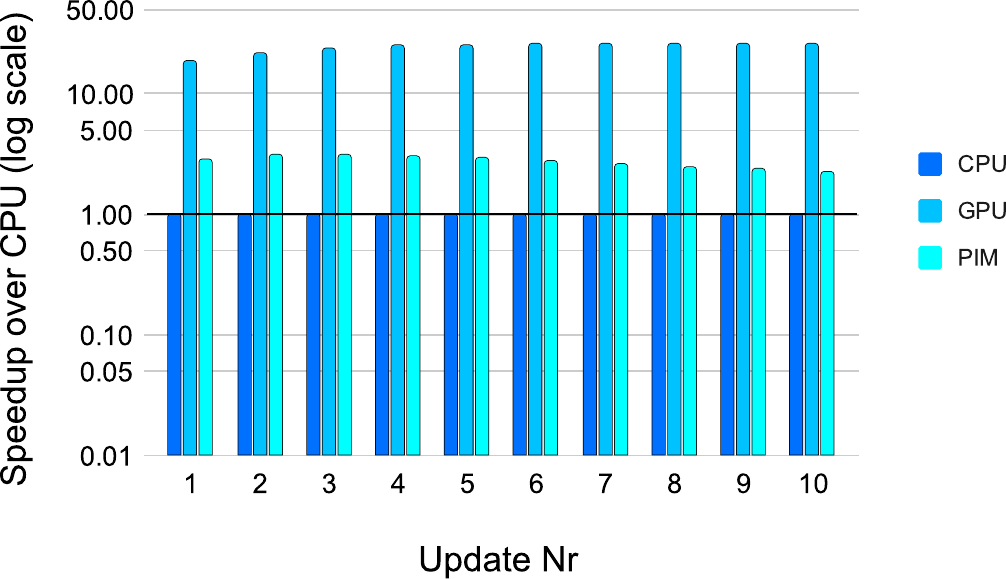}
    \caption{Speedup of PIM and GPU implementations over the baseline CPU implementation when counting the exact triangle number with multiple graph updates (cumulative time).}
\label{fig:speedupDynamic}
\end{figure}

Figure \ref{fig:speedupDynamic} illustrates that in a dynamic environment, the ability to handle and process COO-formatted graphs directly results in a significant performance improvement. The CPU implementation, limited by its internal need for a CSR formatted graph, must perform a complete conversion of the entire graph—including all updates to that point—before starting the triangle counting process every iteration. In contrast, the GPU and PIM implementations can directly update their internal graph representations and quickly begin counting the triangles formed by the newly updated set of edges, requiring a lower cumulative time to process all the updates. 

\section{Conclusions}

In this work we present the first implementation of triangle counting in a real-world Processing-In-Memory architecture. We implement different strategies to overcome the constraints of current PIM architectures. We utilize vertex coloring to avoid costly cross-PIM core communications. We leverage reservoir sampling at the PIM core level to address limitations imposed by the constrained DRAM banks in these cores. We make use of uniform sampling, which aims to reduce the volume of transfers between the host processor and PIM cores. We also address the challenge of high-degree nodes in graphs when using an edge-iterator approach for triangle counting by using the Misra-Gries summary.
\par
In the evaluation of our triangle counting algorithm designed for a real-world PIM system, we conduct extensive analyses for varying critical parameters. We demonstrate how adjusting the number of colors utilized for vertex coloring, varying the parameters used by the Misra-Gries summary, and discarding edges in uniform sampling affect the execution time for different graphs. Finally, we compare our PIM TC implementation against CPU and GPU implementations. Although our performance is lower than that of a GPU, it is important to note that GPU technology has benefited from decades of development, whereas PIM technology is relatively new. Nevertheless, we show a significant speedup of the PIM system over a CPU implementation in our comparison of triangle counting on different architectures when considering dynamic graphs.

\section*{Acknowledgments}
This work was supported in part by the Big-Mobility project under the Uni-Impresa call of the University of Padova, by the MUR
PRIN 2022TS4Y3N EXPAND project, and by MUR PNRR CN00000013
National Center for HPC, Big Data and Quantum Computing. We acknowledge the generous gifts from our industrial partners, including Google, Huawei, Intel, and Microsoft. This work is supported in part by the ETH Future Computing Laboratory (EFCL), Huawei ZRC Storage Team, Semiconductor Research Corporation, AI Chip Center for Emerging Smart Systems (ACCESS), sponsored by InnoHK funding, Hong Kong SAR, and the European Union’s Horizon program for research and innovation [101047160 - BioPIM].

\printbibliography

\end{document}